\documentclass[showkeys]{revtex4}
\usepackage{graphicx,amsmath,amsfonts}
\begin{document}

\title{EXTENDED SIEGERT THEOREM IN THE RELATIVISTIC INVESTIGATION OF THE DEUTERON PHOTODISINTEGRATION REACTION}

\author{S.G. Bondarenko}
\email{bondarenko@jinr.ru}
\author{V.V. Burov}
\email{burov@theor.jinr.ru}
\affiliation{BLTP, Joint Institute for Nuclear Research, 141980 Dubna, Russia}
\author{K.Yu. Kazakov}
\email{kazakov.konst@gmail.com}
\author{D.V. Shulga}
\email{denis@iput.dvgu.ru}
\affiliation{LTNP, Far Eastern National University, Sukhanov Str. 8, 690095 Vladivostok, Russia}

\begin{abstract}

The contribution of the two-body exchange current is investigated for the reaction of the deuteron photodisintegration in the framework of the Bethe-Salpeter formalism and with using extended Siegert theorem. This theorem allow to express the reaction amplitude in terms of extended electric and magnetic dipole moments of the system. The resultant analytical expression is faultless with respect to both translation and gauge invariance. It permits to perform calculations of the deuteron photodisintegration cross section and polarization observables taking into account two-body exchange current implicitly. 
\end{abstract}


\maketitle

\section{Introduction}

One of the most important method of the nuclen-nucleon interaction investigation is the reaction of the electrons and photons scattering on the deuteron. 
In the previous works~\cite{KaShi},~\cite{KaShu},~\cite{BoBuKaShu} the reaction of deuteron photodisintegration has been studied on basis of the Bethe-Salpeter formalism within the framework of one-body approximation.
Final state interaction has been considered besides.
However it was insufficient for complete description of experimental data in the wide range of energy.
With increasing of photon energy the significant contribution to differential cross section comes from meson exchange currents.
An appropriate method of description of such effects is the extended Siegert theorem~\cite{Siegert},~\cite{extended}.
Let's consider it briefly.

\section{The extended Siegert's theorem}

Generalization of Siegert's theorem is made without conventional
decomposition of the EM current into two parts -- the convection
current associated with the motion of nucleus as a whole and the
intrinsic current~\cite{Shebeko}.

First on needs to construct the matrix element for the absorption
of a real photon with three-momentum $\textbf{q}$, while the
nuclear system makes a transition from an internal state
$|\textbf{P}_i\: i\rangle$ to a final internal state
$|\textbf{P}_f=\textbf{P}_i+\textbf{q}\: f\rangle$.

The required matrix element is given by
\begin{equation}
    S_{if}^\gamma=
        \int\!d^4 x~ \langle\textbf{P}_f\:f|
        j_\mu(x)
        |\textbf{P}_i\:i \rangle\,
        \langle 0_\gamma|A^\mu(x)|1_\gamma \rangle.
    \label{SMatrixInitial}
\end{equation}

Applying the invariance with respect to four translation  we may
write
\begin{equation}
    j_\mu(x) = \textrm{e}^{\imath P\cdot x} j_\mu(0) \textrm{e}^{-\imath P\cdot x},
    \label{CurrentTranslated}
\end{equation}
where $P$ is the operator of total four momentum of the system,
i.e.
\begin{equation}
    P | \textbf{P}_i\: i \rangle= P_i | \textbf{P}_i\: i \rangle,
    \quad
    P | \textbf{P}_f\: f \rangle = P_f | \textbf{P}_f\: f \rangle.
    \label{States}
\end{equation}
Introducing Eq.~\eqref{CurrentTranslated} into
Eq.~\eqref{SMatrixInitial} we find
\begin{equation}
    S_{if}^\gamma=
        \langle\textbf{P}_f\:f |
        j_\mu(0)~
        |\textbf{P}_i\: i \rangle
        \langle 0_\gamma | A^\mu(P_f-P_i)
        | 1_\gamma \rangle.
    \label{SMatrixWorking}
\end{equation}
with
\begin{equation}
    A^\mu(q) = \dfrac{1}{\sqrt{2\omega}}\sum\limits_{s=0,1,2,3}
    \varepsilon^\mu_s(\textbf{q})\bigl(a_s(\textbf{q})+a_s^+(\textbf{q})\bigr)
    \:\textrm{e}^{\imath \omega t},
    \label{AFourierTransformation}
\end{equation}
where $a_s(\textbf{q})$ and $a_s^+(\textbf{q})$ are the
destruction and creation operators, respectively, for a photon of
three-momentum $\textbf{q}$ and unit four-polarization vector
$\varepsilon$. The transversality condition is $\varepsilon\cdot
q=0$.

Now we will separate the matrix element into two parts 
using the identity
\begin{equation}
    \boldsymbol{\varepsilon}\textrm{e}^{-\imath \textbf{q}\cdot \textbf{x}} =
    \int\limits_0^1 \!d\lambda
    \left\{\boldsymbol{\nabla}_{\textbf{x}}\bigl(
        \boldsymbol{\varepsilon}\cdot \textbf{x}\:\textrm{e}^{-\imath \lambda
        \textbf{q}\cdot\textbf{x}}\bigr)
        +\imath\omega\lambda \:\textbf{x} \times \boldsymbol{\varepsilon}'
    \textrm{e}^{-\imath \lambda
    \textbf{q}\cdot\textbf{x}}\right\},
    \label{Identity}
\end{equation}
where $\boldsymbol{\varepsilon}$ and
$\boldsymbol{\varepsilon}'=\textbf{q} \times
\boldsymbol{\varepsilon}/\omega$ represent the unit electric and
magnetic polarization vectors, respectively. This identity is
equivalent to a gauge transformation of the EM potential in QED
\begin{align}\label{}
A_\mu^{\textrm{T}}(x)\to
A_\mu(x)=A_\mu^{\textrm{T}}(x)-\partial_\mu \Lambda(x),
\end{align}
where the superscript $T$ on $A$ indicates it is in the transverse
gauge, i.e.
$A^{\textrm{T}}=(0,\boldsymbol{\varepsilon}^{\textrm{T}}\textrm{e}^{-\imath
\textbf{q}\cdot\textbf{x}})$, and $\Lambda(x)$ is such an function
of $x$ that the current has the Siegert limit. Suppressing the
time dependence it means that
\begin{equation}\label{}
\lim\limits_{\omega\to0}\Lambda(\textbf{x})=\boldsymbol{\varepsilon}\cdot\textbf{x}.
\end{equation}
Thus the identity~\eqref{Identity} is a gauge transformation with
the following choice for $\Lambda(\textbf{x})$ (Foldy
gauge~\cite{Foldy})
\begin{equation}\label{}
\Lambda(\textbf{x})=\boldsymbol{\varepsilon}\cdot\textbf{x}\int\limits_0^1\!d\lambda\:
\textrm{e}^{-\imath \lambda \textbf{q}\cdot\textbf{x}}\cong
\boldsymbol{\varepsilon}\cdot\textbf{x}\bigl(1-\tfrac12\imath
\textbf{q}\cdot\textbf{x}-\tfrac16(\textbf{q}\cdot\textbf{x})^2+\textit{O}(\textbf{q}^3)\bigr).
\end{equation}
In principle, it should make no difference for the final result at
low energies. However, in approximate many-body calculations with
$j_\mu$ not necessarily conserved, the choice of gauge does become
important. The Foldy gauge has the good theoretical property that
$\textbf{A}(\textbf{x})$ projects out from
$\textbf{j}(\textbf{x})$ only the magnetic part, i.e. all magnetic
effects are contained in $\textbf{j}_{\textrm{m}}(\textbf{x})$. It
means that the knowledge of the nonrelativistic one-body nucleon
charge density $\rho_{(1)}$ is optimized in the Foldy gauge.

Substituting Eq.~\eqref{Identity} into Eq.~\eqref{SMatrixWorking},
one has
\begin{equation}
    \langle\textbf{P}_f\: f|
    j_\mu(0)| \textbf{P}_i\: i \rangle
    =
    \imath(E_f-E_i) \textbf{D}_{if}(\textbf{q}) - \imath\textbf{q} \times
    \textbf{M}_{if}(\textbf{q}),
    \label{CurrentElementWithDM}
\end{equation}
where $\textbf{q}= \textbf{P}_f - \textbf{P}_i$, $E_i$ and $E_f$
is the total energy of the nuclear system in the initial and final
states. The quantities $\textbf{D}_{fi}(\textbf{q})$ and
$\textbf{M}_{fi}(\textbf{q})$ has the form
\begin{eqnarray}
    \textbf{D}_{if}(\textbf{q})&=& \int\limits_0^1 \!d\lambda \int \!d\textbf{x}~
    \textbf{x}~ \rho_{if}(\textbf{x}; \textbf{P}_i)~ \textrm{e}^{\imath \lambda \textbf{q}\cdot \textbf{x}}, \label{D} \\
    \textbf{M}_{if}(\textbf{q})&=& \int\limits_0^1 \! d\lambda \int \!d\textbf{x}~
    \lambda ~\left[\textbf{x}\times \textbf{j}_{if}(\textbf{x}; \textbf{P}_i)\right]~ \textrm{e}^{\imath \lambda \textbf{q}
    \cdot \textbf{x}}. \label{M}
\end{eqnarray}
with
\begin{eqnarray}
    \langle\textbf{P}_f\: f |
    \textbf{j}(0)|\textbf{P}_i\: i \rangle&=& \int d\textbf{x}~\textrm{e}^{\imath (\textbf{P}_f - \textbf{P}_i)\cdot\textbf{x}}
    \textbf{j}_{if}(\textbf{x}; \textbf{P}_i) , \label{Jfi} \\
    \langle\textbf{P}_f\: f |\rho(0)
    |\textbf{P}_i\: i \rangle&=& \int d\textbf{x}~\textrm{e}^{\imath (\textbf{P}_f - \textbf{P}_i)\cdot\textbf{x}}
    \rho_{if}(\textbf{x}; \textbf{P}_i). \label{Rhofi}
\end{eqnarray}
Reversing Eqs.~\eqref{Rhofi} and \eqref{Jfi}, we verify that
\begin{eqnarray}
    \textbf{j}_{if}(\textbf{x}; \textbf{P}_i) &=& \frac{1}{(2 \pi)^3}\int\! d\textbf{p}~\textrm{e}^{-\imath \textbf{p}\cdot \textbf{x}}
    \,
    \langle\textbf{p} + \textbf{P}_i \: f|
    \textbf{j}(0)|\textbf{P}_i\: i \rangle,
    \label{InverseJfi} \\
    \rho_{if}(\textbf{x}; \textbf{P}_i) &=& \frac{1}{(2 \pi)^3}
    \int d\textbf{p}~\textrm{e}^{-\imath \textbf{p}\cdot
    \textbf{x}}\,
    \langle\textbf{p} + \textbf{P}_i \: f|
    \rho(0)|\textbf{P}_i\: i \rangle. \label{InverseRhofi}
\end{eqnarray}
From Eqs.~\eqref{D} and \eqref{Rhofi} we immediately find
\begin{equation}
    \imath \textbf{q} \cdot\textbf{D}_{if}(\textbf{q})
    =
    \langle\textbf{P}_f\: f|\rho(0)| \textbf{P}_i\:i \rangle
    -
    \langle\textbf{P}_i\: f|
    \rho(0)
    | \textbf{P}_i\:i \rangle.
    \label{kD}
\end{equation}
At this point we are ready to write the $S$-matrix element in
terms of the matrix elements operators $\textbf{D}(\textbf{q})$
and $\textbf{M}(\textbf{q})$. Substituting
Eq.~(\ref{CurrentElementWithDM}) into Eq.~\eqref{SMatrixWorking}
and using the relation \eqref{kD} we obtain
\begin{eqnarray}
    S_{if}^\gamma &=&
    \langle \textbf{P}_i\: f |
    \rho(0)| \textbf{P}_i\: i \rangle\, \langle 0_\gamma|\phi(q)|1_\gamma\rangle
    + \imath\left[ \textbf{q}\: \langle 0_\gamma|\phi(q)|1_\gamma\rangle - \omega\:
    \langle 0_\gamma|\textbf{A}(q)|1_\gamma\rangle\right]\cdot \textbf{D}_{if}(\textbf{q})\nonumber \\
    &+& \imath\left[ \textbf{q} \times \textbf{M}_{if}(\textbf{q})\right]\cdot\langle 0_\gamma|\textbf{A}(q)|1_\gamma\rangle.
    \label{SMatrixWithDMandA}
\end{eqnarray}

Finally introducing the strength of the electric and magnetic
fields (in momentum space)
\begin{eqnarray}
    \textbf{E}(q)&=& \imath \omega\textbf{A}(q) - \imath \textbf{q}~ \phi(q), \label{EImpRepr}\\
    \textbf{H}(q)&=& \imath \textbf{q} \times \textbf{A}(q), \label{HImpRepr}
\end{eqnarray}
we cast the $S$-matrix element in the manifestly  gauge
independent form
\begin{eqnarray}
    S_{i\gamma\to f}  =
    - \langle 0_\gamma|\textbf{E}(q)|1_\gamma\rangle\cdot \textbf{D}_{if}(\textbf{q})- \langle 0_\gamma|\textbf{H}(q)|1_\gamma\rangle
    \cdot  \textbf{M}_{if}(\textbf{q})
    +\langle\textbf{P}_i\: f|
    \rho(0)| \textbf{P}_i\: i \rangle \langle 0_\gamma|\phi(q)|1_\gamma\rangle.
    \label{SMatrixWithDM}
\end{eqnarray}

The scattering amplitude for the photon absorption is written in
gauge independent form
\begin{equation}\label{}
T_{if}^\gamma= - \langle
0_\gamma|\textbf{E}(0)|1_\gamma\rangle\cdot
\textbf{D}_{fi}(\textbf{q})- \langle
0_\gamma|\textbf{H}(0)|1_\gamma\rangle
    \cdot  \textbf{M}_{fi}(\textbf{q})
\end{equation}
with
\begin{eqnarray}
\langle 0_\gamma|\textbf{E}(0)|1_\gamma\rangle & =&
\dfrac{1}{\sqrt{2\omega}}\:\bigl(\omega\boldsymbol{\varepsilon}(\textbf{q})-
\textbf{q}\varepsilon_0(\textbf{q})\bigr),\\
\langle 0_\gamma|\textbf{H}(0)|1_\gamma\rangle & =&
\dfrac{1}{\sqrt{2\omega}}\:\textbf{q}\times\boldsymbol{\varepsilon}(\textbf{q}).
\end{eqnarray}

The last term in Eq.~\eqref{SMatrixWithDM} is equal to zero, since
it is proportional to the matrix element of the total charge
between orthogonal states. This equation is consistent with
requirements of gauge invariance as well as invariance with
respect to translations.

\section{Application to relativistic formalism}

Now we shall write down $T$-matrix in Siegert's form within
Bethe-Salpeter formalism~\eqref{SMatrixWithDM}
\begin{eqnarray}
    T_{if}^\gamma &=&
    - \dfrac{1}{\sqrt{2\omega}}\:\bigl(\omega\boldsymbol{\varepsilon}(\textbf{q})-
\textbf{q}\varepsilon_0(\textbf{q})\bigr)\cdot\textbf{D}_{if}(\textbf{q})-
\dfrac{1}{\sqrt{2\omega}}\:[\textbf{q}\times\boldsymbol{\varepsilon}(\textbf{q})]\cdot\textbf{M}_{if}
    (\textbf{q}).
\end{eqnarray}
The operators $\textbf{D}_{fi}(\textbf{q})$ and
$\textbf{M}_{fi}(\textbf{q})$ could be obtained from
Eqs.~\eqref{D} and \eqref{M} with the use of
Eqs.~\eqref{InverseJfi} and \eqref{InverseRhofi}
\begin{eqnarray}
    \textbf{D}_{if}(\textbf{q}) &=& -\imath \int\limits_0^1 \dfrac{d\lambda
    }{\lambda}\,
    \nabla_{\textbf{q}} \langle\lambda \textbf{q} +\textbf{P}_i f |
    \rho(0) | \textbf{P}_i i \rangle, \label{D through matrix element} \\
    \textbf{M}_{if}(\textbf{q}) &=& -\imath \int\limits_0^1
    d\lambda\,
    \nabla_{\textbf{q}} \times \langle\lambda \textbf{q} +\textbf{P}_i f |
    \textbf{j}(0)| \textbf{P}_i i \rangle. \label{M through matrix element}
\end{eqnarray}

Substituting expression for the matrix element of the EM current between
two relativistic states

\begin{eqnarray}
    \langle\textbf{P}_f f | j^\mu(0)| \textbf{P}_i i \rangle
    &=& \imath \int \dfrac{ds dk}{(2\pi)^8} ~\bar \chi_{P_f}(s)~
    \Lambda^\mu(s, k; P_f, P_i)~ \chi_{P_i}(k),
    \label{CurrentMatrixElement in BS}
\end{eqnarray}

into Eqs.~\eqref{D through matrix element} and \eqref{M through matrix
element} we obtain
\begin{eqnarray}
    \textbf{D}_{if}(\textbf{q}) &=& \int\limits_0^1 \dfrac{d\lambda
    }{\lambda}\,
    \nabla_{\textbf{q}} \int \frac{d^4s d^4s'}{(2\pi)^8} ~\bar \chi_{P_i+\lambda q}(s)~
    \Lambda^0(s, s'; P_i+\lambda k, P_i)~ \chi_{P_i}(s'), \label{D in BS} \\
    \textbf{M}_{if}(\textbf{q}) &=& \int\limits_0^1 d\lambda\,
    \nabla_{\textbf{q}} \times \int \frac{d^4s d^4s'}{(2\pi)^8} ~\bar \chi_{P_i+\lambda q}(s)~
    \boldsymbol{\Lambda}(s, s'; P_i+\lambda q, P_i)~ \chi_{P_i}(s').
    \label{M in BS}
\end{eqnarray}
Using the transverse gauge in calculations, i.e.
$\varepsilon_0^{\textrm{T}}=0$ and
$\boldsymbol{\varepsilon}^{\textrm{T}}\cdot\textbf{q}=0$, one
finds that
\begin{eqnarray}
    T_{if}^\gamma &=&
    - \sqrt{\dfrac{\omega}{2}}\:\boldsymbol{\varepsilon}_\rho^{\textrm{T}}\cdot\textbf{D}_{if}(\textbf{q})-
\dfrac{1}{\sqrt{2}}\:\boldsymbol{\varepsilon}^{\textrm{T}'}_\rho\cdot\textbf{M}_{if}
    (\textbf{q}),\qquad \rho=\pm1.
\end{eqnarray}
One concludes that this matrix elements gives response of the
nuclear system in two transverse perpendicular directions (defined
by the three-vectors $\boldsymbol{\varepsilon}$ and
$\boldsymbol{\varepsilon}'$) with respect to the photon
three-momentum $\textbf{q}$.

\section{Results and discussions}

\begin{figure}[h]
\includegraphics[height=75mm]{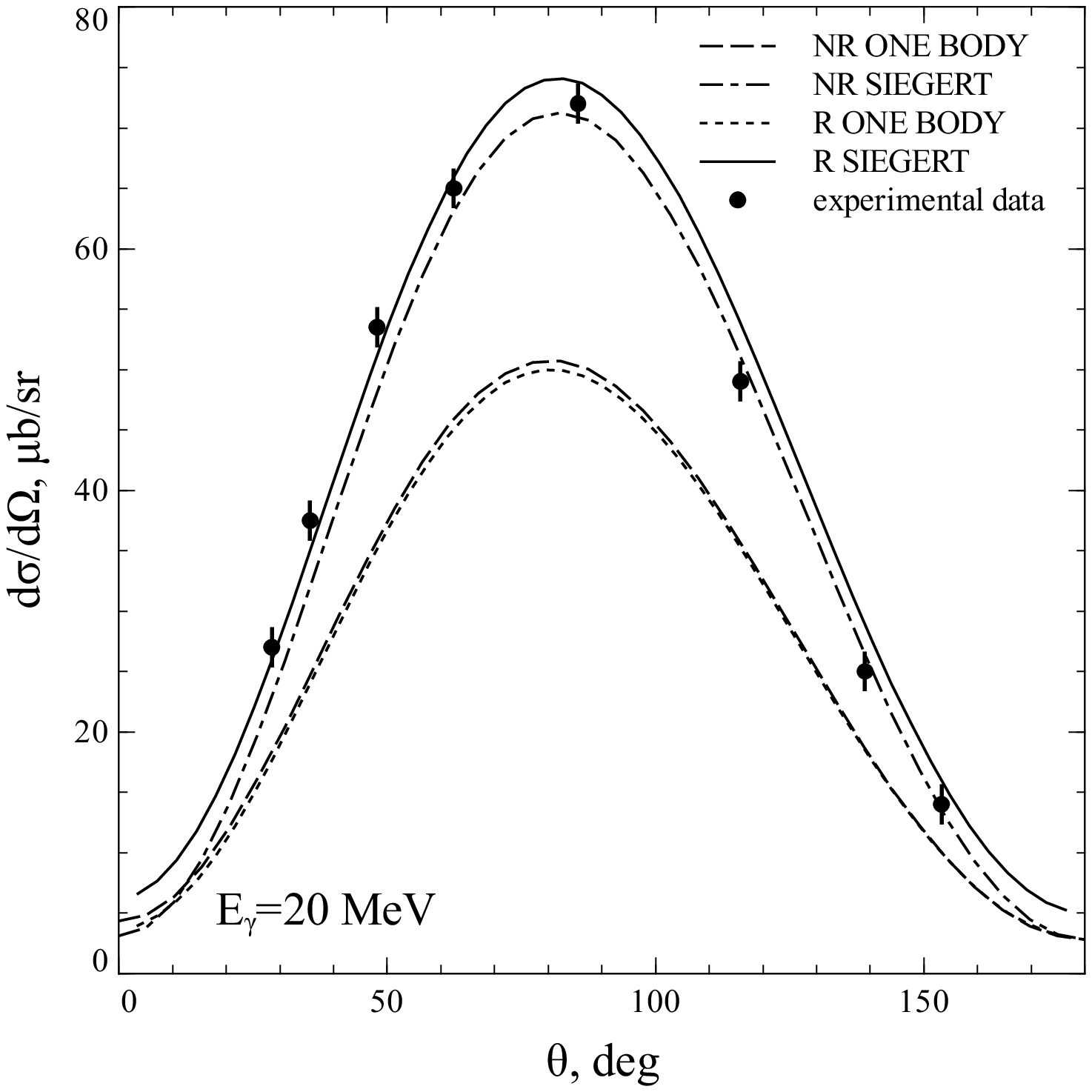}
\includegraphics[height=75mm]{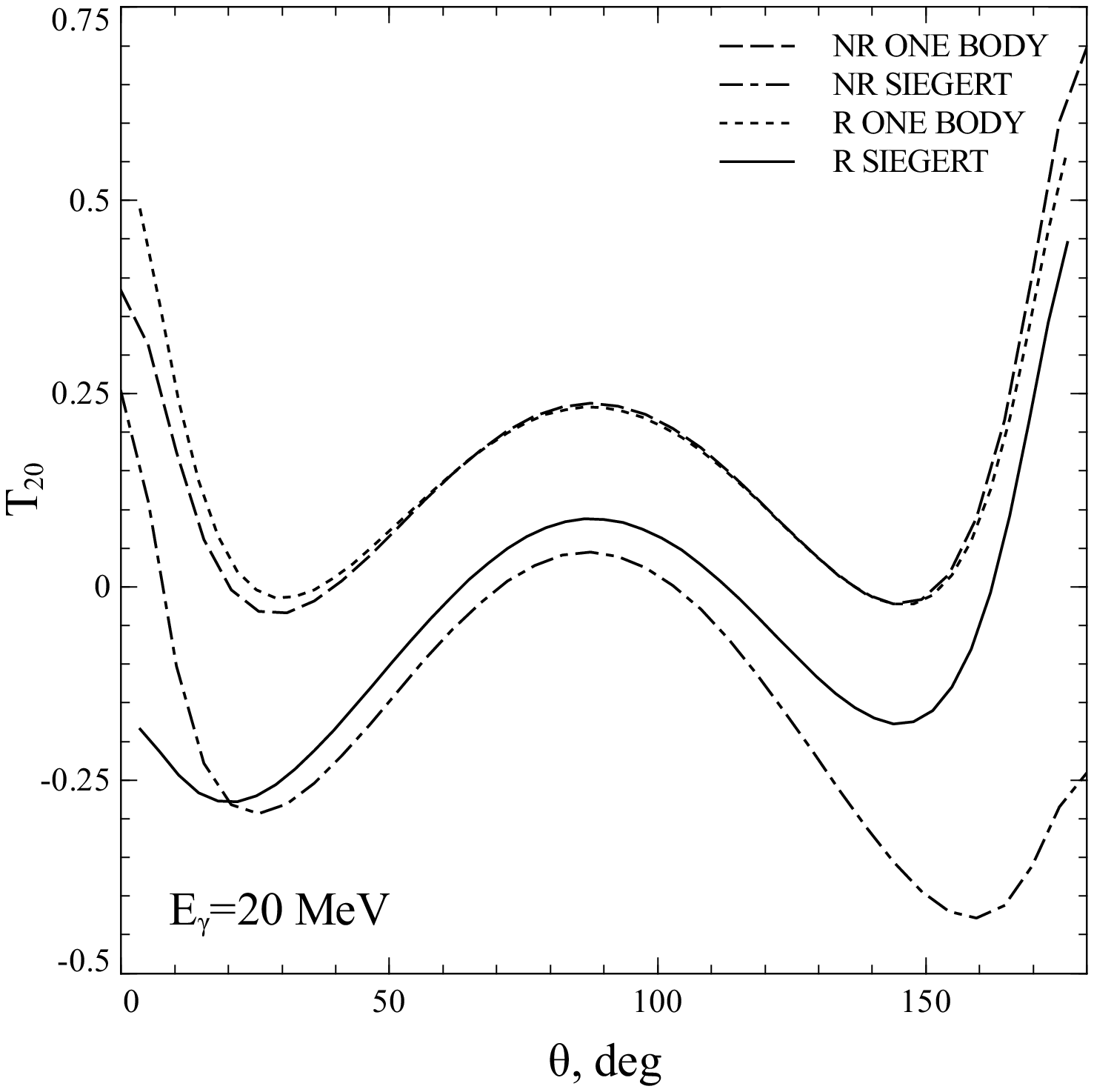}
\includegraphics[height=75mm]{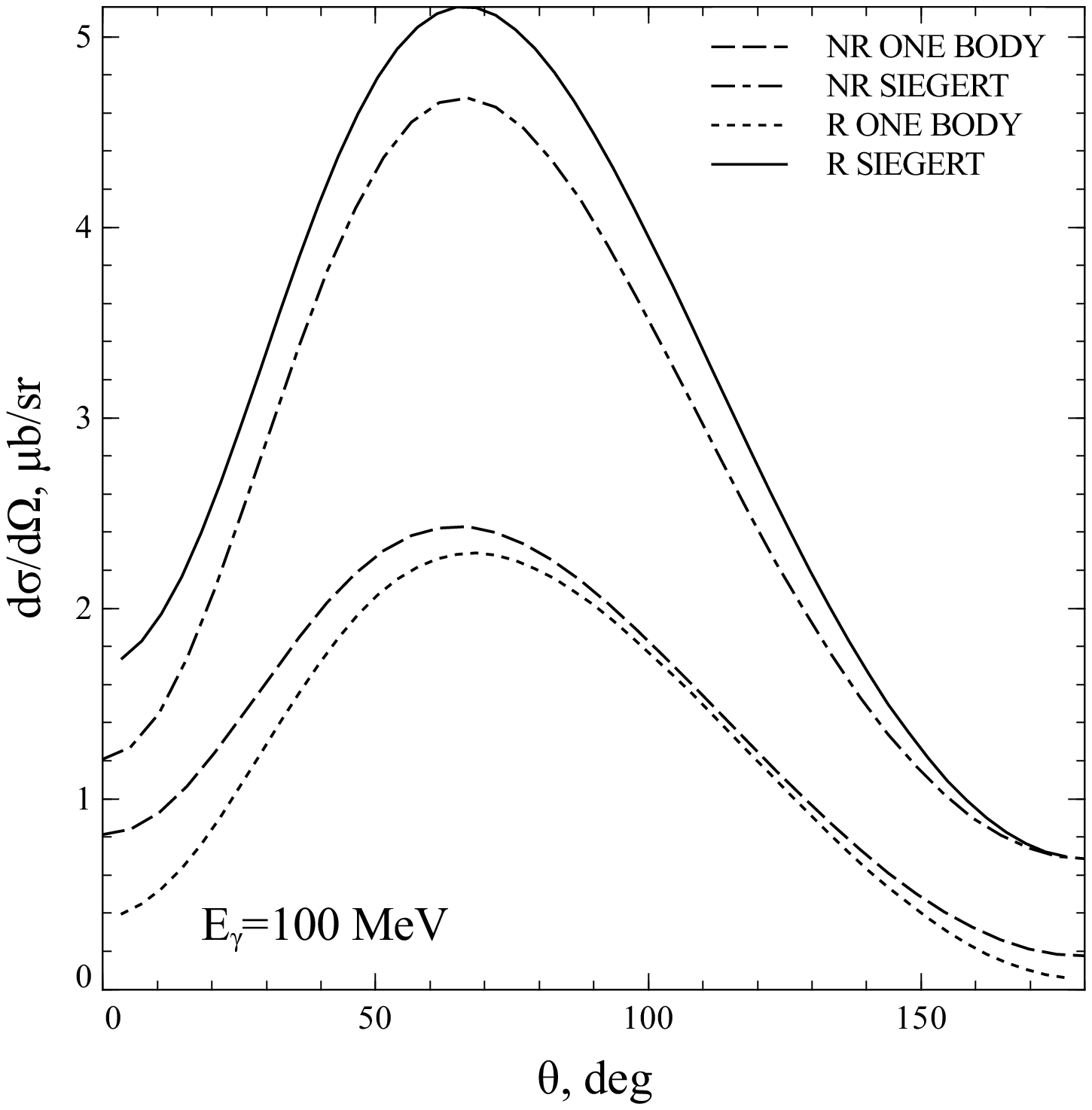}
\includegraphics[height=75mm]{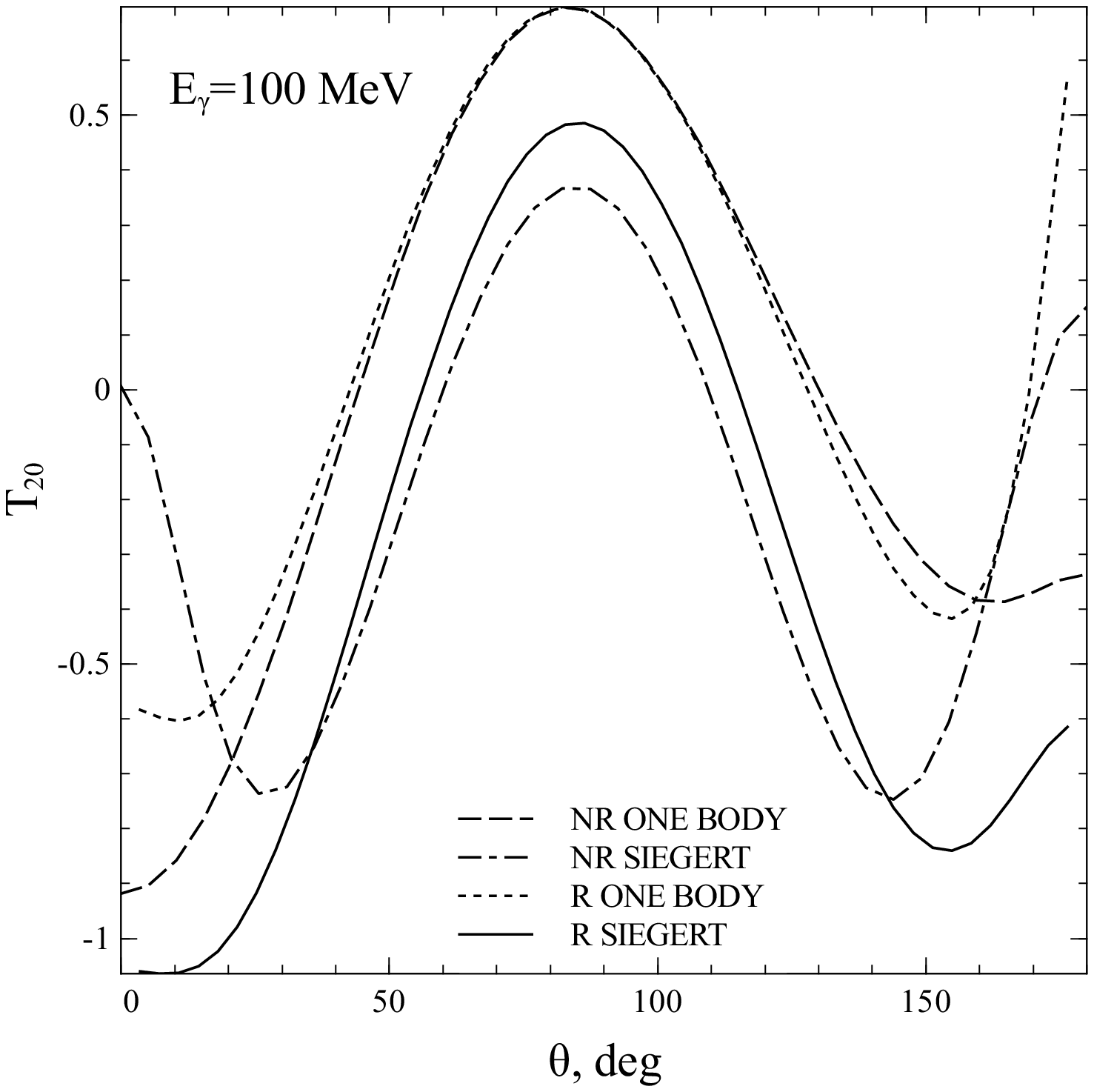}
\includegraphics[height=75mm]{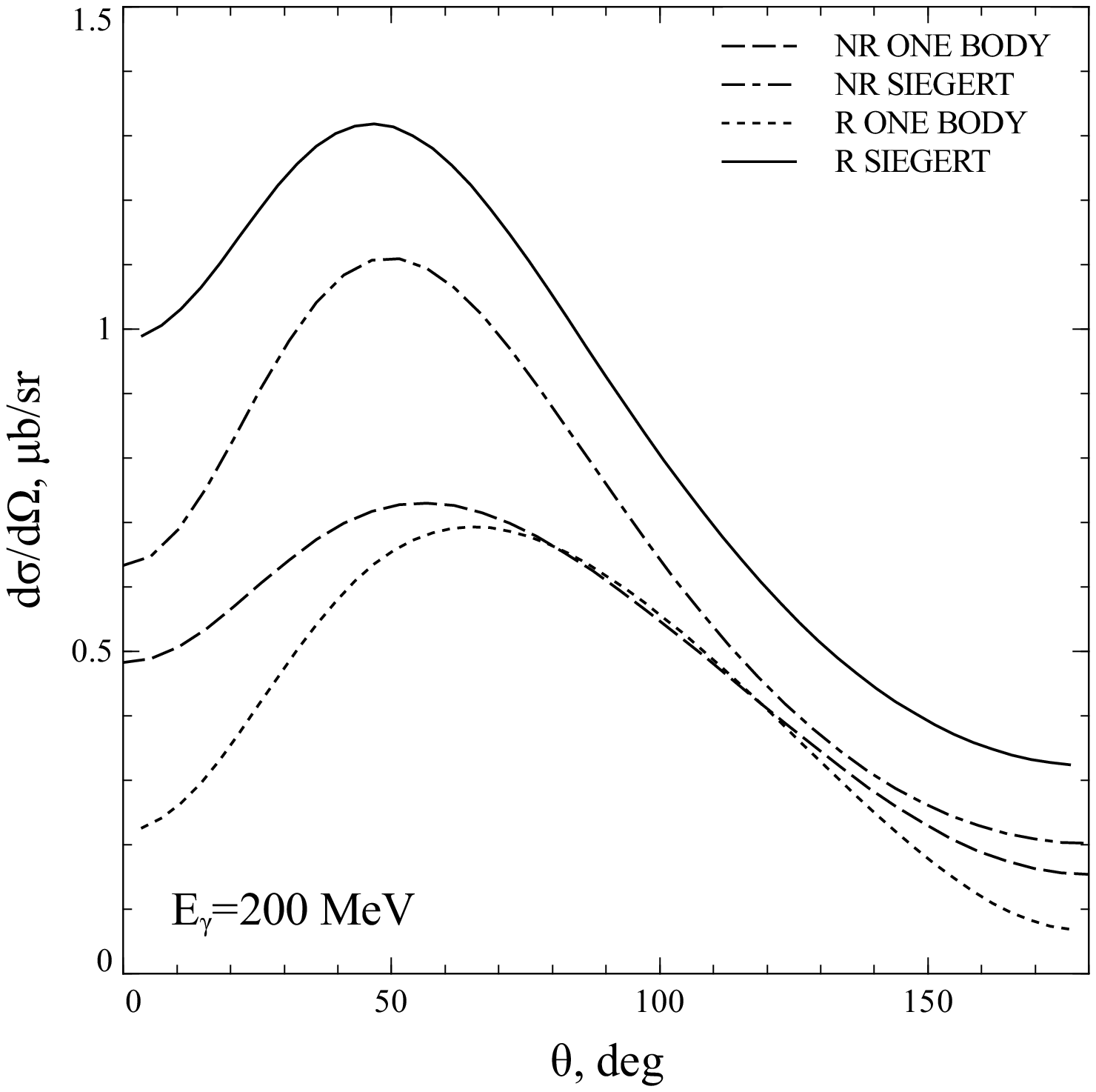}
\includegraphics[height=75mm]{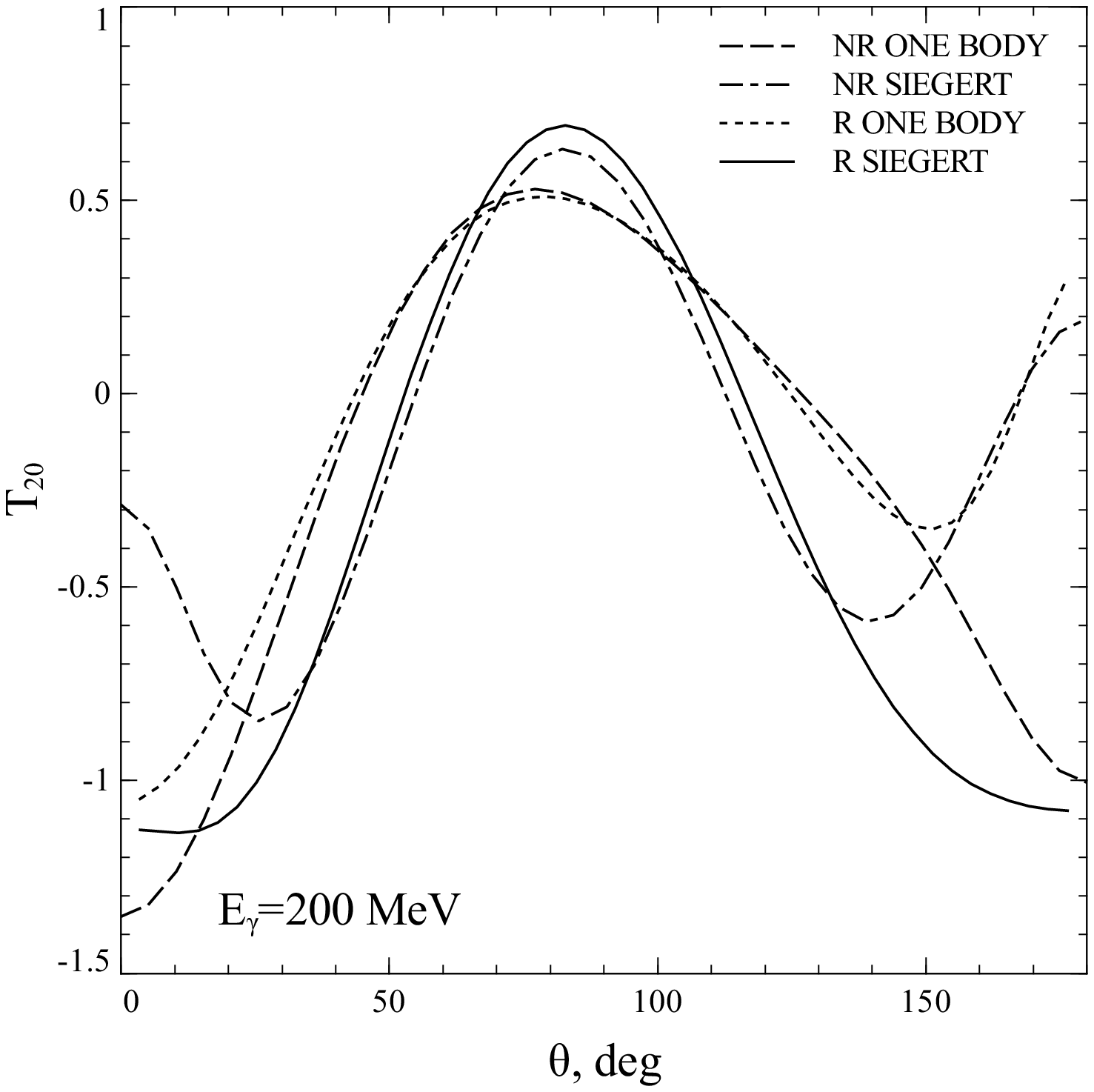}
\caption{Differential cross section and tensor target asymmetry $T_{20}$. Notation of the curves: nonrelativistic one-body current (long-dashed); nonrelativistic two-body current (dash-dotted); relativistic one-body current (dotted); relativistic two-body current (full). Experimental data is taken from~\cite{arenhovel}.}
\label{Fig1}
\end{figure}

\begin{figure}[h]
\includegraphics[height=75mm]{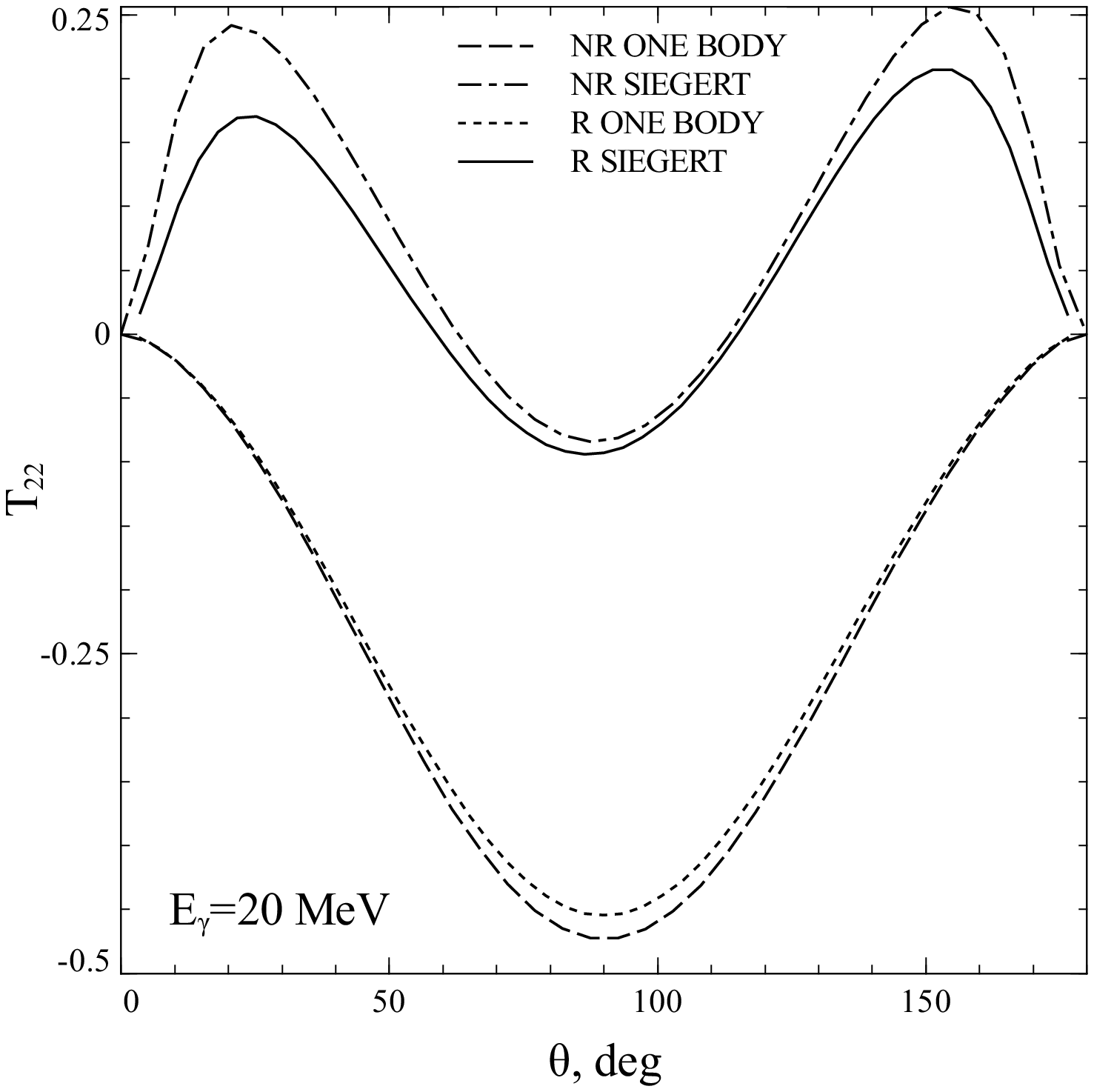}
\includegraphics[height=75mm]{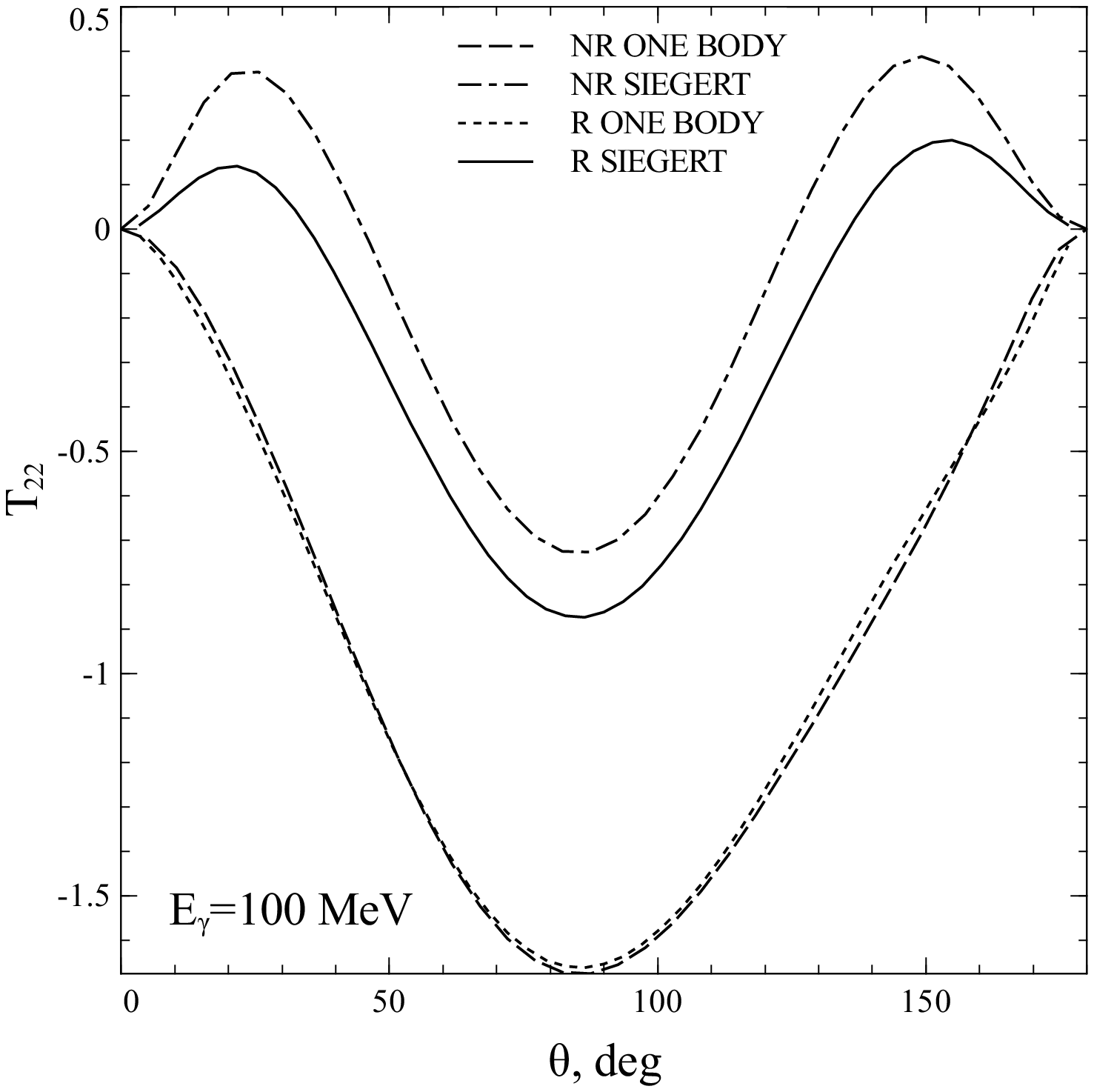}
\includegraphics[height=75mm]{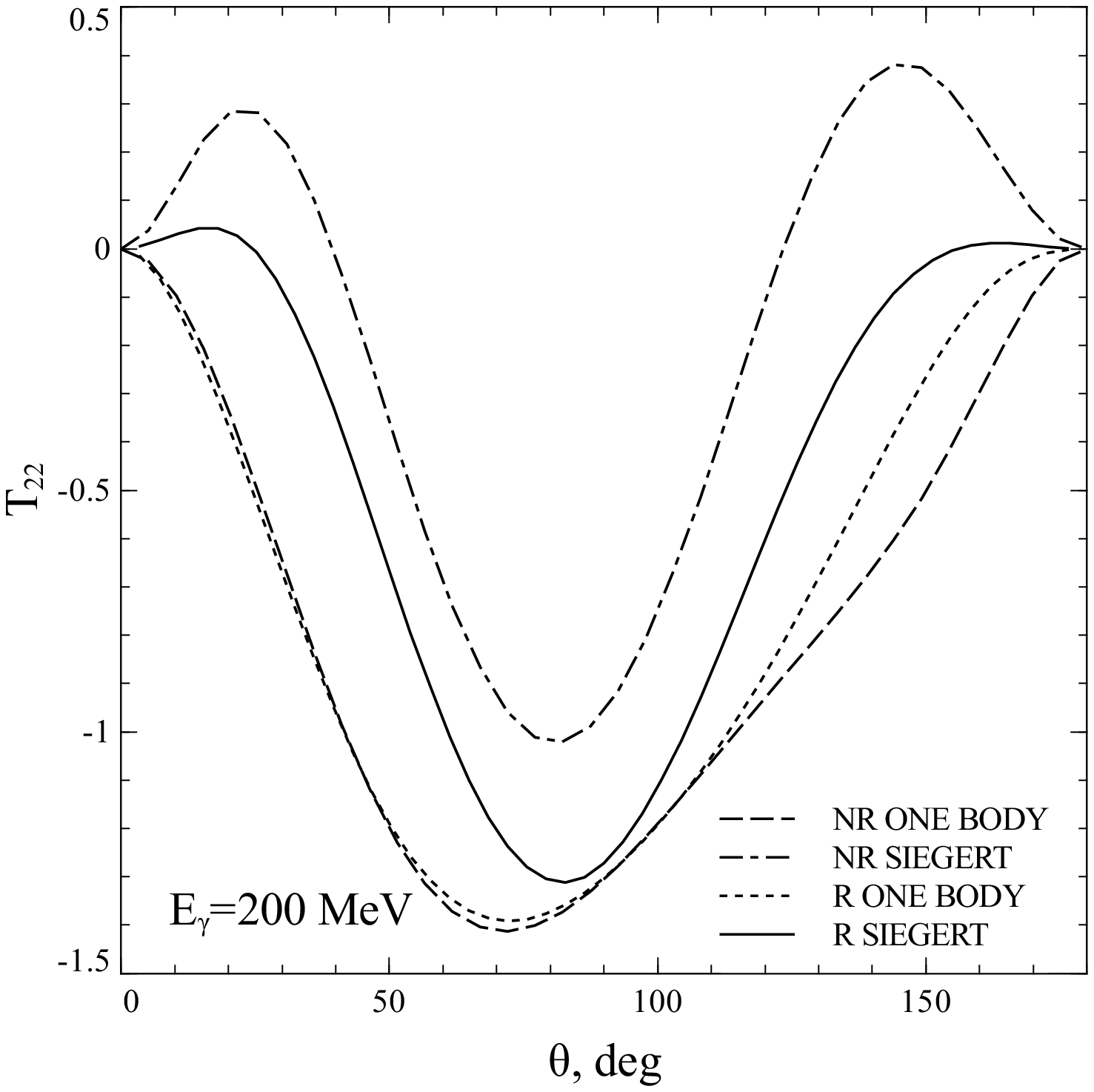}
\includegraphics[height=75mm]{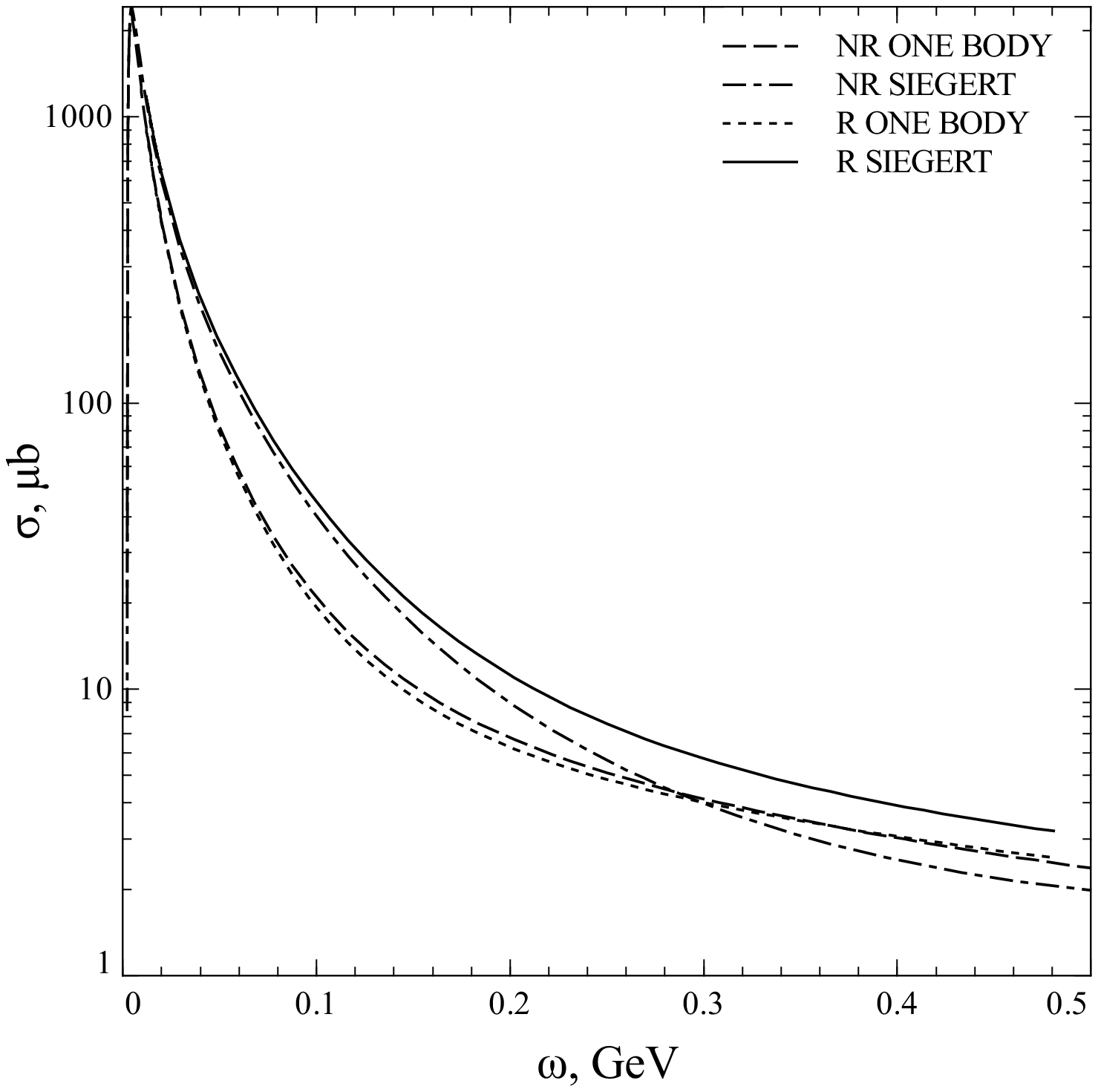}
\caption{Tensor target asymmetry $T_{22}$ and total cross section. Notation of the curves: nonrelativistic one-body current (long-dashed); nonrelativistic two-body current (dash-dotted); relativistic one-body current (dotted); relativistic two-body current (full).}
\label{Fig2}
\end{figure}

The results of our calculations are depicted in the Fig.~\ref{Fig1} and Fig.~\ref{Fig2}.
The calculations has performed in nonrelativistic (Shrodinger equation) and relativistic (Bethe-Salpeter equation) models.
In both cases we use Graz II potential of nucleon-nucleon interaction and carry out the computations with and without including two-body current effectively via extended Siegert theorem. We haven't taken into account final state interaction in this work.
Notations are explained in the figure's captions.

It is seen in the first plot of Fig.~\ref{Fig1} that two-body effects give a large contribution to the differential cross section even at the photon energy equal to 20 MeV. One-body approximation is 30\% less than experimental data. Calculations with using extended Siegert theorem allow to agree with experimental data.

Relativistic effects give larger contribution to two-body current than to one body current. We can see that practically for all the plots at the figures, particularly for $T_{22}$ at 200 MeV. For $T_{20}$ we can see large contribution from relativistic effects even at photon energy equal to 20 Mev.

\end{document}